\documentclass[preprint,12pt]{elsarticle}  
\usepackage{graphicx} 
\usepackage{amssymb} 
\usepackage{amsmath} 

\journal{Physics Letters B} 

\begin{document} 

\begin{frontmatter} 

\title{Few-body calculations of $\eta$-nuclear quasibound states} 
\author{N.~Barnea} 
\author{E.~Friedman} 
\author{A.~Gal\corref{cor1}} 
\cortext[cor1]{corresponding author: Avraham Gal, avragal@savion.huji.ac.il}   
\address{Racah Institute of Physics, The Hebrew University, 91904 
Jerusalem, Israel} 

\begin{abstract} 
We report on precise hyperspherical-basis calculations of $\eta NN$ and 
$\eta NNN$ quasibound states, using energy dependent $\eta N$ interaction 
potentials derived from coupled-channel models of the $S_{11}$ 
$N^{\ast}(1535)$ nucleon resonance. The $\eta N$ attraction generated in 
these models is too weak to generate a two-body bound state. No $\eta NN$ 
bound-state solution was found in our calculations in models where 
Re~$a_{\eta N}\lesssim 1$~fm, with $a_{\eta N}$ the $\eta N$ scattering 
length, covering thereby the majority of $N^{\ast}(1535)$ resonance models. 
A near-threshold $\eta NNN$ bound-state solution, with $\eta$ separation 
energy of less than 1~MeV and width of about 15~MeV, was obtained in the 
2005 Green-Wycech model where Re~$a_{\eta N}\approx 1$~fm. The role of 
handling self consistently the subthreshold $\eta N$ interaction is carefully 
studied. 
\end{abstract} 


\end{frontmatter}

\section{Introduction} 
\label{sec:intro} 

The $\eta N$ interaction has been studied extensively in photon- and 
hadron-induced production experiments on free and quasi-free nucleons, 
and on nuclei \cite{KW15}. These experiments suggest that the near-threshold 
$\eta N$ interaction is attractive, but are unable to quantify this 
statement in any precise manner. Nevertheless, $\eta$ production data 
on nuclei provide some useful hints on possible $\eta$ quasibound states 
for very light species where, according to Krusche and Wilkin (KW) 
``the most straightforward (but not unique) interpretation of the data 
is that the $\eta d$ system is unbound, the $\eta ^4$He is bound, but that 
the $\eta ^3$He case is ambiguous" \cite{KW15}. Indeed, the prevailing 
theoretical consensus since the beginning of the 2000s, based on $\eta NN$ 
Faddeev calculations, is that $\eta d$ quasibound or resonance states are 
ruled out for acceptable $\eta N$ interaction strengths \cite{Deloff00,GP00}. 
Instead, the $\eta d$ system may admit virtual states \cite{FA00,WG01,Gar03}. 
Searching for reliable few-body calculations of the $A=3,4$ $\eta$-nuclear 
systems, we are aware of none for $\eta NNNN$ and of only one $\eta NNN$ 
Faddeev-Yakubovsky calculation \cite{FA02}, although not sufficiently 
realistic, that finds no $\eta ^3$H quasibound state. Rigorous few-body 
calculations substantiating the KW conjecture quoted above are therefore 
called for. The present work fills some of this gap, reporting precise 
calculations of $\eta NN$ and of $\eta NNN$ few-body systems using the 
hyperspherical basis methodology \cite{LO13}, similarly to the calculations 
reported in Ref.~\cite{BGL12} for the $\bar K NN$ and $\bar K NNN$ systems. 
Particular attention is given in the present work to the subthreshold energy 
dependence of the $\eta N$ interaction in a way not explored before in $\eta$ 
few-body calculations. 

\begin{figure}[htb] 
\begin{center} 
\includegraphics[width=0.48\textwidth]{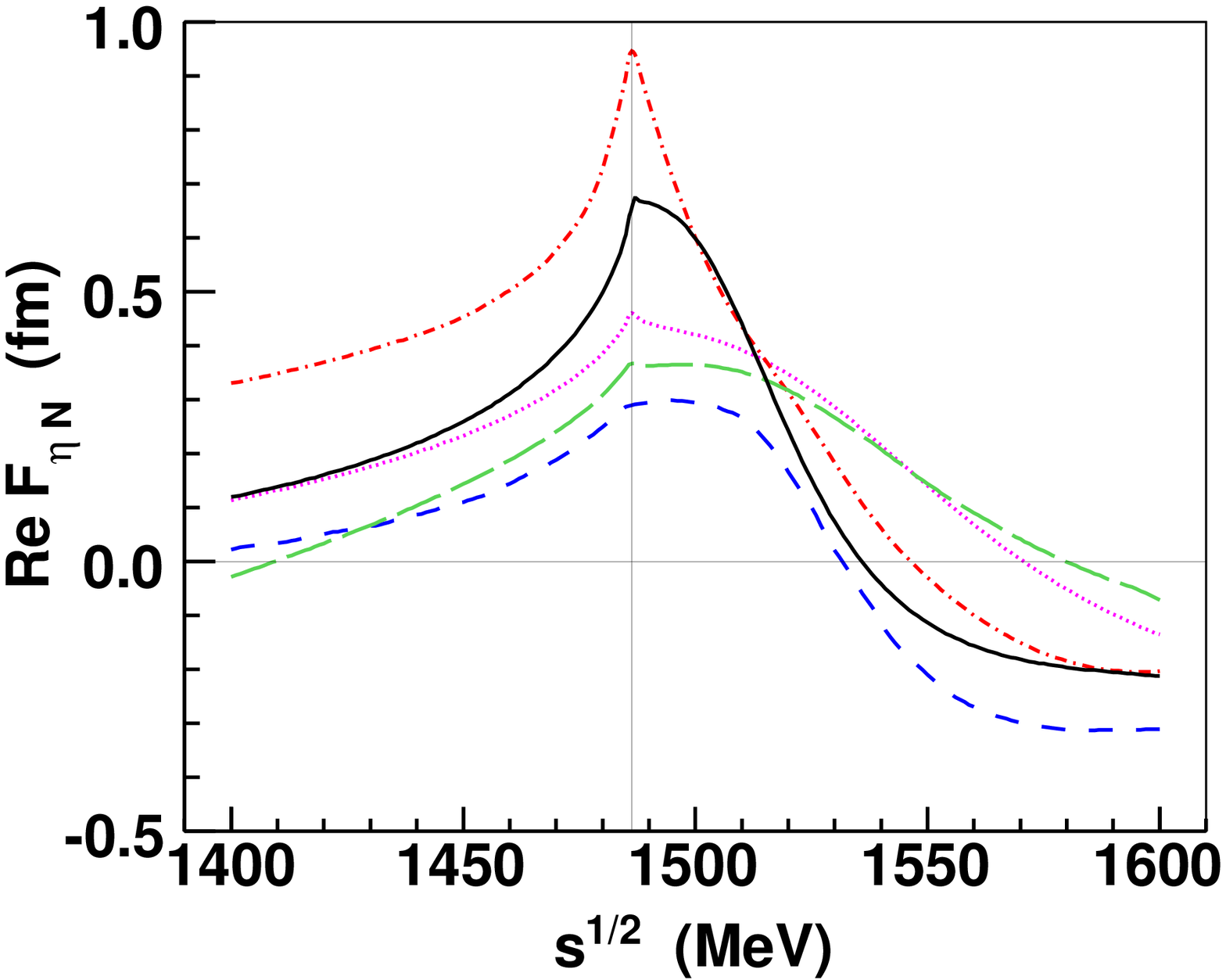} 
\includegraphics[width=0.48\textwidth]{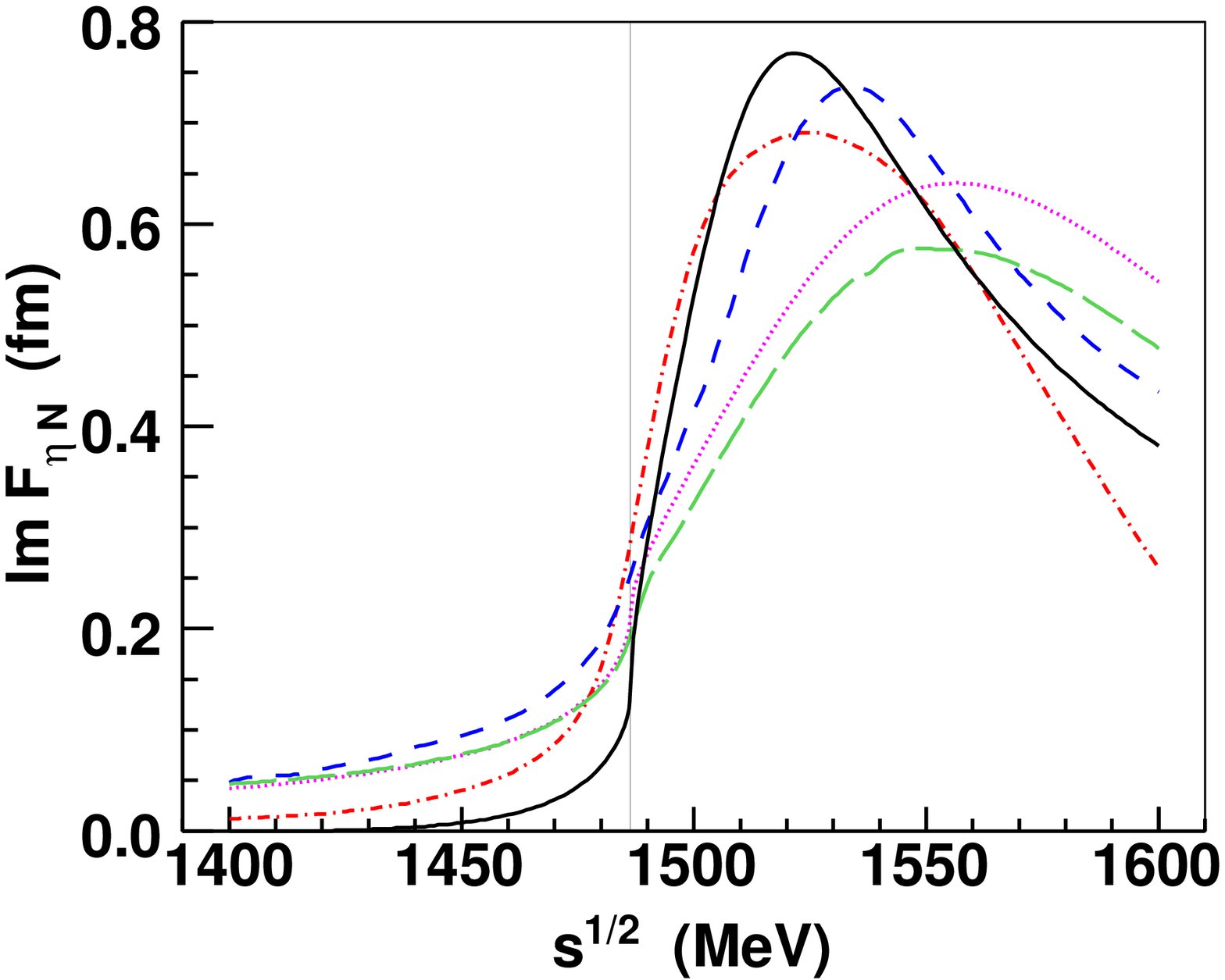} 
\caption{Real (left panel) and imaginary (right panel) parts of the $\eta N$ 
cm scattering amplitude $F_{\eta N}(\sqrt{s})$ as a function of the total cm 
energy $\sqrt{s}$ in five meson-baryon coupled-channel interaction models, 
in decreasing order of Re$\;a_{\eta N}$. Dot-dashed curves: GW \cite{GW05}; 
solid: CS \cite{CS13}; dotted: KSW \cite{KSW95}; long-dashed: M2 \cite{MBM12}; 
short-dashed: IOV \cite{IOV02}. The thin vertical line denotes the $\eta N$ 
threshold. Figure adapted from \cite{GFBCMG14}. } 
\label{fig:aEtaN1.eps} 
\end{center} 
\end{figure} 

Theoretically, the $\eta N$ interaction has been studied in coupled-channel 
models that seek to fit or, furthermore, generate dynamically the prominent 
$N^{\ast}(1535)$ resonance which peaks about 50~MeV above the $\eta N$ 
threshold. Such models result in a wide range of values for the real part 
of the $\eta N$ scattering length $a_{\eta N}$, from 0.2~fm \cite{KWW97} 
to almost 1.0~fm \cite{GW05}. Most of these analyses constrain the imaginary 
part Im~$a_{\eta N}$ within a considerably narrower range of values, from 
0.2 to 0.3~fm. This is demonstrated in Fig.~\ref{fig:aEtaN1.eps} where the 
real and imaginary parts of the $\eta N$ center-of-mass (cm) scattering 
amplitude $F_{\eta N}(\sqrt{s})$ are plotted as a function of the cm energy 
$\sqrt{s}$ for several coupled channel models. The $\eta N$ threshold, where 
$F_{\eta N}(\sqrt{s_{\rm th}})=a_{\eta N}$, is denoted by a thin vertical 
line. We note that both real and imaginary parts of $F_{\eta N}(\sqrt{s})$ 
below threshold decrease monotonically in all of these models upon going 
deeper into the subthreshold region, displaying however substantial model 
dependence. This will become important for the $\eta$ few-body calculations 
reported here. 

Beginning with the pioneering work by Haider and Liu \cite{HL86}, 
and using input values of $a_{\eta N}$ within these specified ranges, 
several $\eta$-nucleus optical-model bound-state calculations concluded 
that $\eta$ mesons are likely to bind in sufficiently heavy nuclei, 
certainly in $^{12}$C and beyond \cite{HL02,GR02,JNH02,FGM13,CFGM14}. 
In the few-body calculations reported here we find no $\eta d$ quasibound 
states for values of Re~$a_{\eta N}$ as large as about 1~fm. We do find, 
however, a very weakly bound and broad $\eta ^3$H--$\eta ^3$He isodoublet 
pair for Re~$a_{\eta N}\approx 1$~fm by solving the $\eta NNN$ four-body 
problem. 

The paper is organized as follows. In section~\ref{sec:v} we construct local 
energy-dependent single-channel potentials $v_{\eta N}$ that reproduce 
two of the $s$-wave scattering amplitudes $F_{\eta N}(\sqrt{s})$ shown 
in Fig.~\ref{fig:aEtaN1.eps}, GW \cite{GW05} and CS \cite{CS13}. 
In section~\ref{sec:KLM} we sketch the hyperspherical-basis formulation 
and solution of the $\eta NN$ and $\eta NNN$ Schroedinger equations 
below threshold using these derived $\eta N$ potentials and realistic 
energy-independent $NN$ potentials. Because of the substantial energy 
dependence of $v_{\eta N}$ in the subthreshold region, a self consistency 
requirement \cite{BGL12} is applied so that the input energy argument 
of the two-body potential $v_{\eta N}$ for convergent few-body 
solutions is consistently related to some energy expectation 
values in the resulting quasibound state. Results are presented 
and discussed in section~\ref{sec:res}, followed by a brief summary 
and outlook section~\ref{sec:concl}.

\section{Construction of $\eta N$ effective potentials} 
\label{sec:v} 

We seek to construct energy-dependent local $\eta N$ potentials $v_{\eta N}$ 
that reproduce the $\eta N$ scattering amplitude $F_{\eta N}(\sqrt{s})$ 
below threshold in given models, e.g. from among those shown in 
Fig.~\ref{fig:aEtaN1.eps}. For convenience, the energy argument $E$ 
introduced in this section is defined with respect to the $\eta N$ threshold, 
$E\equiv\sqrt{s}-\sqrt{s_{\rm th}}$, and should not be confused with the 
binding energy of the $\eta NN$ and $\eta NNN$ few-body states studied in 
subsequent sections. 

We define $v_{\eta N}$ in the form 
\begin{equation} 
v_{\eta N}(E;r)=-\frac{4\pi}{2\mu_{\eta N}}\,b(E)\,\rho_{\Lambda}(r), \;\;\;\; 
(\hbar=c=1)
\label{eq:v(E)} 
\end{equation} 
with $\mu_{\eta N}$ the reduced $\eta N$ mass and where $\rho_{\Lambda}$ is 
a Gaussian normalized to 1: 
\begin{equation} 
\rho_{\Lambda}(r)= \left(\frac{\Lambda}{2\sqrt{\pi}}\right)^3\, 
\exp \left(-\frac{\Lambda^2 r^2}{4}\right). 
\label{eq:rho} 
\end{equation} 
$\Lambda$ is a scale parameter, inversely proportional to the range of 
$v_{\eta N}$. Its physically admissible values are discussed in subsection 
\ref{subsec:scale} below. Two representative values are used here, $\Lambda$=2 
and 4~fm$^{-1}$. For a given value of $\Lambda$, one needs to determine the 
energy-dependent strength parameter $b(E)$ of $v_{\eta N}$, as described in 
the following subsection \ref{subsec:strength}.

\subsection{Solution} 
\label{subsec:strength} 
  
\begin{figure}[thb] 
\begin{center}  
\includegraphics[width=0.46\textwidth]{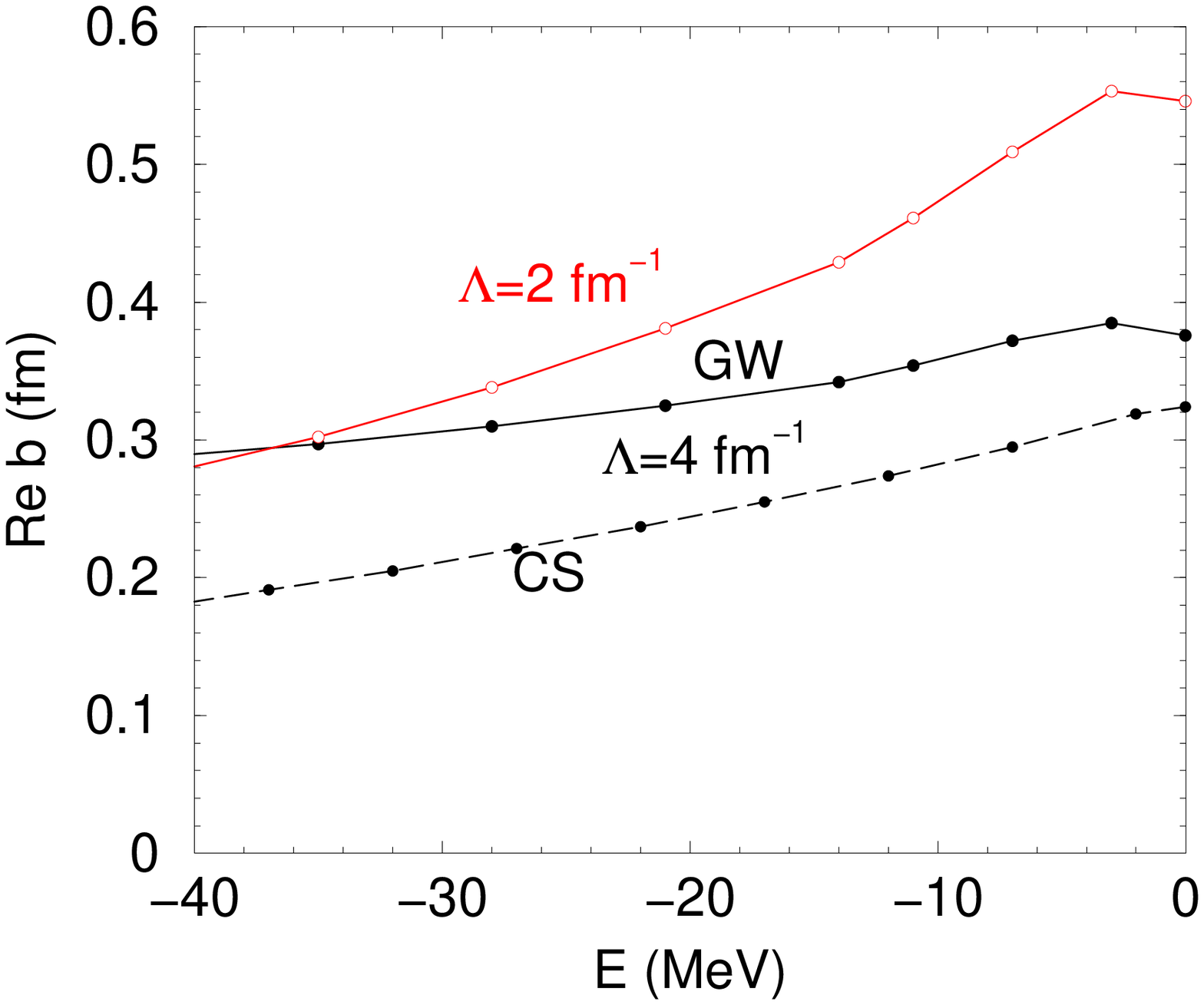} 
\includegraphics[width=0.48\textwidth]{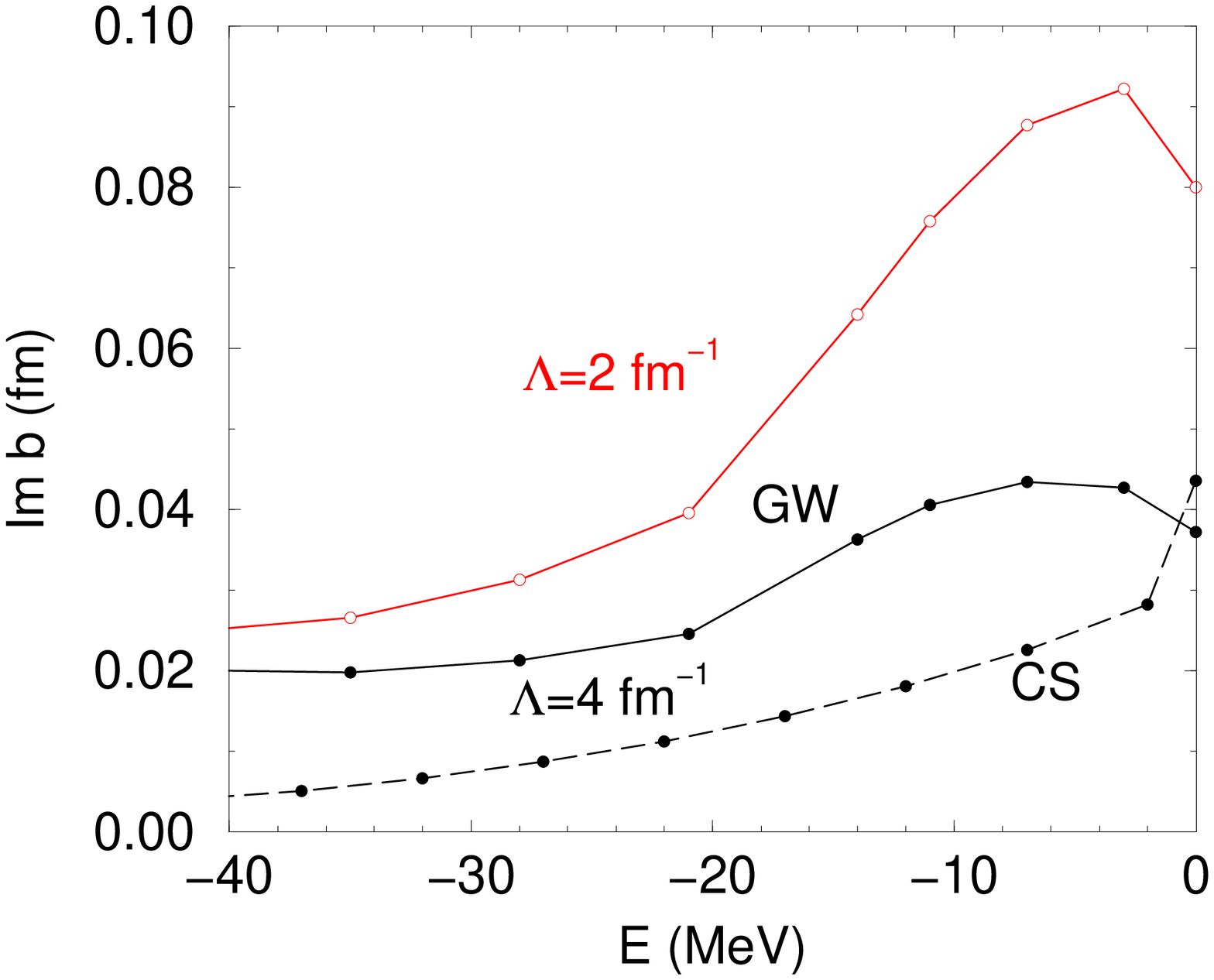} 
\caption{Real (left panel) and imaginary (right panel) parts of the strength 
parameter $b(E)$ of the $\eta N$ effective potential (\ref{eq:v(E)}), 
for subthreshold energies $E<0$, obtained from the scattering amplitudes 
$F_{\eta N}^{\rm GW}$ \cite{GW05} and $F_{\eta N}^{\rm CS}$ \cite{CS13} shown 
in Fig.~\ref{fig:aEtaN1.eps}. Two choices of the scale parameter $\Lambda$ 
are made for GW, both resulting in the same $F_{\eta N}^{\rm GW}(E)$, 
and just one for CS.} 
\label{fig:Wycfit4.eps} 
\end{center} 
\end{figure}

Given a specific value of the scale parameter $\Lambda$, the two-body $s$-wave 
Schroedinger equation 
\begin{equation} 
-\frac{1}{2\mu_{\eta N}}u''(r)+v_{\eta N}(E;r)u(r)=Eu(r) 
\label{eq:Schroedinger} 
\end{equation} 
is solved for energies above ($E>0$) and below ($E<0$) threshold. 
The radial wavefunction $u(r)$ satisfies the boundary conditions 
\begin{equation} 
u(r=0)=0, \;\;\;\;\;\;  u(r\to\infty)\propto r(\cos\delta_0\,j_0(kr) - 
\sin\delta_0\,n_0(kr)), 
\label{eq:boundary} 
\end{equation} 
where $k=\sqrt{2\mu_{\eta N}E}$, $j_0$ and $n_0$ are spherical Bessel and 
Neumann functions, respectively, and $\delta_0(E)$ is the complex $s$-wave 
phase shift derived by imposing these boundary conditions on the wave-equation 
solution. Above threshold, the wave number $k$ is real and taken positive. 
Below threshold, $k=i\kappa$ with $\kappa >0$. The scattering amplitude $F$ 
is then given by 
\begin{equation} 
F_{\eta N}(E)=\frac{1}{k(\cot\delta_0-i)}. 
\label{eq:delta} 
\end{equation} 
This procedure was used in Ref.~\cite{HW08} for constructing effective 
$\bar K N$ potentials below threshold. In the present case, the subthreshold 
values of the complex strength parameter $b(E)$ in Eq.~(\ref{eq:v(E)}) were 
fitted to the complex phase shifts $\delta (E)$ derived from subthreshold 
scattering amplitudes $F_{\eta N}(E)$ in several of the coupled-channel 
models of Fig.~\ref{fig:aEtaN1.eps}. This is shown for the GW~\cite{GW05} 
and CS~\cite{CS13} models in Fig.~\ref{fig:Wycfit4.eps}, using two values 
of the scale parameter $\Lambda$=2 and 4~fm$^{-1}$ for GW and just one 
value $\Lambda=4$~fm$^{-1}$ for CS. The curves $b(E)$ are seen to decrease 
monotonically in going deeper below threshold, except for small kinks near 
threshold that reflect the threshold cusp of Re~$F_{\eta N}(E=0)$ in 
Fig.~\ref{fig:aEtaN1.eps}. Comparing models GW and CS for the {\it same} 
scale parameter $\Lambda=4$~fm$^{-1}$, one observes larger values of $b(E)$ 
in model GW than in CS, for both real and imaginary parts below threshold, 
in line with the larger GW subthreshold amplitudes compared with the 
corresponding CS amplitudes. We note furthermore that Im~$b(E)$$\ll$Re~$b(E)$ 
in both models by almost an order of magnitude, 
see Fig.~\ref{fig:Wycfit4.eps}, which justifies treating Im~$v_{\eta N}$ 
perturbatively in the applications presented below. 

\begin{figure}[htb] 
\begin{center} 
\includegraphics[width=0.48\textwidth]{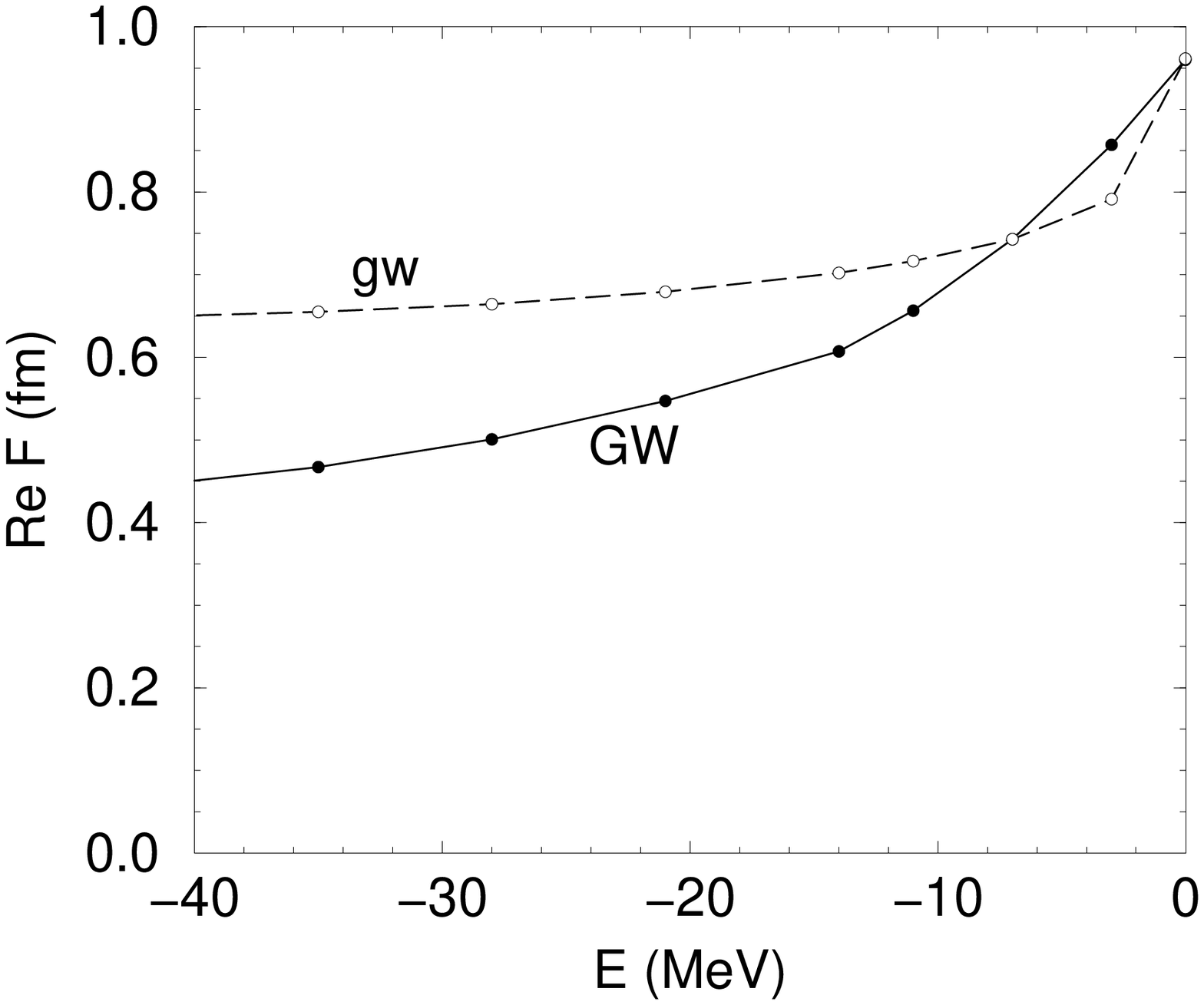} 
\includegraphics[width=0.48\textwidth]{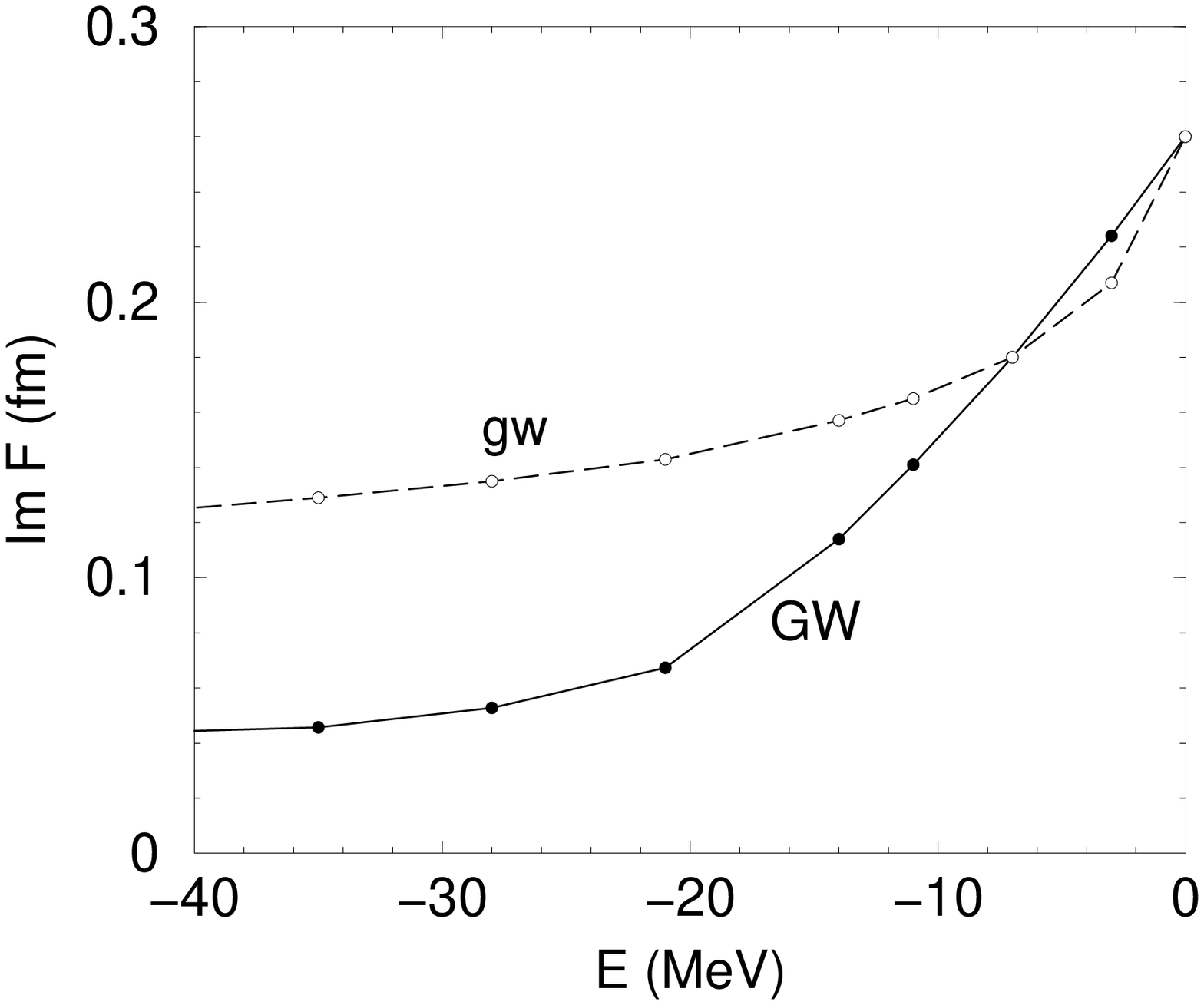} 
\caption{Real (left panel) and imaginary (right panel) parts of the 
subthreshold amplitude $F_{\eta N}^{\rm GW}(E)$ (solid curves marked GW) 
from Fig.~\ref{fig:aEtaN1.eps}, also generated from an energy dependent 
potential $v_{\eta N}^{\rm GW}(E)$ with $\Lambda=4$~fm$^{-1}$, 
Eqs.~(\ref{eq:v(E)},\ref{eq:rho}), compared to the amplitude (dashed curves 
marked gw) generated from $v_{\eta N}^{\rm GW}(E=0)$.} 
\label{fig:Wycfixedb.eps} 
\end{center} 
\end{figure} 

To demonstrate the extent to which the energy dependence of $b(E)$ is 
essential, we compare in Fig.~\ref{fig:Wycfixedb.eps} the GW subthreshold 
amplitude from Fig.~\ref{fig:aEtaN1.eps}, which is also generated here 
using the $b(E)$ potential strength of Fig.~\ref{fig:Wycfit4.eps} for 
$\Lambda=4$~fm$^{-1}$, to the amplitude marked gw which was calculated 
using a fixed threshold value $b(E=0)$. This latter amplitude is seen 
to decrease too slowly beginning about $E\approx -7$~MeV. 
Obviously, an {\it energy-independent} single-channel potential 
$v_{\eta N}$ fails to reproduce the subthreshold energy dependence 
of the GW coupled-channel scattering amplitude $F_{\eta N}^{\rm GW}(E)$.

\subsection{Choice of scale}
\label{subsec:scale}

It is appropriate at this point to address the model dependence introduced 
in $\eta$-nuclear few-body calculations by the choice of the scale parameter 
$\Lambda$ made in constructing $v_{\eta N}$, Eqs.~(\ref{eq:v(E)}) and 
(\ref{eq:rho}). $\Lambda$ is often identified with the momentum cutoff used 
to renormalize divergent loop integrals in on-shell EFT $N^{\ast}(1535)$ 
models \cite{MBM12,IOV02}. In separable-interaction coupled channel models, 
however, the momentum cutoff is replaced by fitted Yamaguchi form factors 
$(q^2+\Lambda^2)^{-1}$ with a momentum-space range parameter $\Lambda$, 
the Fourier transform of which is a Yukawa potential $\exp(-\Lambda r)/r$ 
with r.m.s radius identical to that of the Gaussian potential shape 
(\ref{eq:rho}). Values of $\Lambda$ from three such $N^{\ast}(1535)$ models, 
including the two used in the present work \cite{GW05,CS13}, are listed in 
Table~\ref{tab:Lambda}. 

\begin{table}[htb] 
\begin{center} 
\caption{The $\eta N$ momentum scale parameter $\Lambda$ from several 
$N^{\ast}(1535)$ separable models.} 
\begin{tabular}{cccc} 
\hline 
Ref. & \cite{KWW97,KSW95} & \cite{GW05} & \cite{CS13} \\ 
\hline 
$\Lambda$~(fm$^{-1}$) & 3.9 & 3.2 & 6.6 \\ 
\hline 
\end{tabular} 
\label{tab:Lambda} 
\end{center} 
\end{table} 

Inspection of Table~\ref{tab:Lambda} reveals a broad range of values that 
$\Lambda$ may assume, starting with $\Lambda\approx 3$~fm$^{-1}$. The 
relatively high value in the third column is rather exceptional for 
meson-baryon separable models. Given this broad spectrum of values spanned for 
$\Lambda$, we chose two values $\Lambda=2$ and $\Lambda=4$~fm$^{-1}$ to study 
the model dependence of our $\eta$-nuclear few-body calculations. The higher 
value, $\Lambda=4$~fm$^{-1}$, corresponds to a Gaussian $\exp(-r^2/R^2)$ 
spatial range $R=2/\Lambda=0.5$~fm, a value which is very close to $R=0.47$~fm 
taken from the systematic EFT approach in Ref.~\cite{HW08} and used in our 
$\bar K$-nuclear few-body calculations \cite{BGL12}. As argued there, choosing 
smaller values for $R$, namely larger values than 4~fm$^{-1}$ for $\Lambda$, 
would be inconsistent with staying within a purely hadronic basis.{\footnote
{The effective energy-dependent $\bar K N$ potential $v_{\bar KN}$ 
constructed by Hyodo and Weise \cite{HW08} reproduces the $\bar KN-\pi\Sigma$ 
coupled-channel scattering amplitude which is the one essential for generating 
dynamically the $\Lambda^{\ast}(1405)$ resonance. In that case, the choice 
of $\Lambda$ must ensure that the ${\bar K}^{\ast}N$ channel that couples 
strongly to $\bar KN$ via normal pion exchange is kept outside of the model 
space in which $v_{\bar K N}$ is valid. This argument leads to a choice 
of $\Lambda = p_{\rm min}(\bar K N\to {\bar K}^{\ast} N)=552$~MeV/c or 
2.8~fm$^{-1}$, corresponding to a Gaussian spatial range of $R=0.71$~fm. 
In a somewhat similar reasoning Garzon and Oset \cite{GO15} recently argued 
for extending the EFT description of the $N^{\ast}(1535)$ resonance to include 
the $\rho N$ channel which couples strongly to the already included $\pi N$ 
channel, although not to $\eta N$. Identifying $\Lambda$ with the minimum 
momentum needed to excite the $\pi N$ system to $\rho N$, we obtain 
$\Lambda = p_{\rm min}(\pi N\to \rho N)=586$~MeV/c or 3.0~fm$^{-1}$.}} 

In the Introduction section we loosely identified the strength of the $\eta N$ 
interaction with the size of the real part of its threshold scattering 
amplitude, Re~$a_{\eta N}\lesssim 1$~fm. However, in terms of the interaction 
potentials $v_{\eta N}$ that enter our few-body calculations, a given 
value of Re~$a_{\eta N}$ does not rule out a broad spectrum of spatial 
ranges, or equivalently momentum scale parameters $\Lambda$, as demonstrated 
in Fig.~\ref{fig:Wycfit4.eps}. A model dependence is thereby introduced into 
our few-body calculations, summarized by stating that the larger the $\eta N$ 
scale parameter $\Lambda$ is, the larger is the $\eta$ separation energy, 
provided it is quasibound. This lack of scale invariance hints towards the 
necessity of including three-body forces, as is expected from an EFT point 
of view \cite{BvK02}. Such three-body forces amount to adding a new free 
parameter determined by tuning it to some $\eta$ few-body experimental data.

\section{$\eta$-nuclear hyperspherical-basis formulation and solution} 
\label{sec:KLM} 

The hyperspherical-basis formulation of meson-nuclear few-body calculations 
was initiated in Ref.~\cite{BGL12} for $\bar K$ mesons. Here we sketch 
briefly the necessary transformation from $\bar K$ mesons to $\eta$ mesons. 
The $N$--body wavefunction ($N=3,4$) in our case consists of a sum over 
products of isospin, spin and spatial components, antisymmetrized with respect 
to nucleons. In the spatial sector  translationally invariant basis functions 
are constructed in terms of one hyper-radial coordinate $\rho$ and a set of 
$3N-4$ angular coordinates [$\Omega_N$], substituting for $N-1$ Jacobi 
vectors. The spatial basis functions are of the form  
\begin{equation} 
\Phi_{n,[K]}(\rho,\Omega_N)=R^{N}_{n}(\rho)
{\cal Y}^{N}_{[K]}(\Omega_N), 
\label{eq:HH} 
\end{equation} 
where $R^{N}_{n}(\rho)$ are hyper-radial basis functions expressible in 
terms of Laguerre polynomials and ${\cal Y}^{N}_{[K]}(\Omega_N)$ are 
hyperspherical-harmonics (HH) functions in the angular coordinates $\Omega_N$ 
expressible in terms of spherical harmonics and Jacobi polynomials. Here, the 
symbol $[K]$ stands for a set of angular-momentum quantum numbers, including 
those of ${\hat L}^2$, ${\hat L}_z$ and ${\hat K}^2$, where ${\hat{\bf K}}$ 
is the total grand angular momentum which reduces to the total orbital angular 
momentum for $N=2$. The HH functions ${\cal Y}^{N}_{[K]}$ are eigenfunctions 
of ${\hat K}^2$ with eigenvalues $K(K+3N-5)$, and $\rho^{K}{\cal Y}^{N}_{[K]}$ 
are harmonic polynomials of degree $K$ \cite{LO13}. 

For the $NN$ interaction we used two forms, the (Minnesota) MN central 
potential \cite{Minn77} and the Argonne AV4' potential \cite{WP02} derived 
from the full AV18 potential by suppressing the spin-orbit and tensor 
interactions and readjusting the central spin and isospin dependent 
interactions. In $s$-shell nuclei the AV4' potential provides an excellent 
approximation to AV18 which pseudoscalar mesons, such as the $\eta$ meson, 
are unlikely to spoil, recalling that their nuclear interactions cannot 
induce $S \leftrightarrow D$ mixing beyond that already accounted for 
by the $NN$ interaction.{\footnote{This was demonstrated in $\bar K$ 
nuclear cluster calculations \cite{BGL12}, see the discussion of Table~1 
therein, where the ${\bar K}(NN)_{I=0}$ 4.7~MeV binding energy contribution 
to the full 15.7~MeV binding energy of $({\bar K} NN)_{I=1/2}$ calculated 
using AV4' is short by only 0.2~MeV from that in a comparable 
calculation~\cite{DHW08} using AV18.}} AV4' and MN differ mostly in 
their short-range repulsion which is much stronger in AV4' than in MN. 

For the $\eta N$ interaction we used the energy-dependent local potential 
Re~$v_{\eta N}$ introduced in Sect.~\ref{sec:v}. In order to distinguish the 
energy $E$ of the few-body system from the energy argument of $v_{\eta N}$, 
the latter is replaced by $\delta\sqrt{s}\equiv\sqrt{s}-\sqrt{s_{\rm th}}$ 
from now on. Following Eq.~(5) in \cite{BGL12}, the subthreshold energy 
argument $\delta\sqrt{s}$ of $v_{\eta N}$, is chosen to agree 
self-consistently with 
\begin{equation} 
\langle\delta\sqrt{s}\rangle = -\frac{B}{A}-\xi_{N}\frac{A-1}{A}\langle 
T_{N:N} \rangle -\frac{A-1}{A}B_{\eta}-\xi_{\eta}\left ( \frac{A-1}{A} 
\right )^2 \langle T_{\eta} \rangle \; , 
\label{eq:sqrt{s}} 
\end{equation} 
where $\xi_{N(\eta)}\equiv m_{N(\eta)}/(m_N+m_{\eta})$, $T_{\eta}$ is 
the $\eta$ kinetic energy operator in the total cm frame, $T_{N:N}$ is 
the pairwise $NN$ kinetic energy operator in the $NN$ pair cm frame, 
$B$ is the total binding energy of the $\eta$-nuclear few-body system 
and $B_{\eta}$ is the $\eta$ ``binding energy", 
$B_{\eta}\equiv -E_{\eta} =-\langle\Psi|(H-H_N)|\Psi\rangle$, where $H_N$ 
is the Hamiltonian of the purely nuclear part in its own cm frame and the 
total Hamiltonian $H$ is evaluated in the overall cm frame. In the limit 
$A\gg 1$, Eq.~(\ref{eq:sqrt{s}}) agrees with the nuclear-matter expression 
given in Refs.~\cite{FGM13,CFGM14} for use in calculating $\eta$-nuclear 
quasibound states. It provides a self-consistency cycle in $\eta$-nuclear 
few-body calculations by requiring that the expectation value $\langle\delta
\sqrt{s}\rangle$ derived from the solution of the Schroedinger equation agrees 
with the input value $\delta\sqrt{s}$ used in $v_{\eta N}$. Since each one 
of the four terms on the r.h.s. of (\ref{eq:sqrt{s}}) is negative, the self 
consistent energy shift $\delta\sqrt{s_{\rm s.c.}}$ is necessarily negative, 
with size exceeding a minimum nonzero value obtained from the first two terms 
in the limit of vanishing $\eta$ binding. 

The potential and kinetic energy matrix elements for a given $\eta$-nuclear 
state with global quantum numbers $I,L,S,J^{\pi}$ were evaluated in the 
HH basis. The $NN$ and $\eta N$ interactions specified above conserve 
$I=I_N$, $S=S_N$ and $L$. Since no $L\neq 0$ $\eta$-nuclear states are 
likely to come out particle stable, our calculations are limited to $L=0$. 
The deuteron in this approximation is a purely $^3S_1$ state. Suppressing 
${\rm Im}\,v_{\eta N}$, the g.s. energy $E_{\rm g.s.}$ was calculated 
in a model space spanned by HH basis functions with eigenvalues 
$K\leq K_{\rm max}$. Self-consistent calculations were done for $\sqrt{s}$ 
ranging from the $\eta N$ threshold down to 30 MeV below. Self consistency 
in $\delta\sqrt{s}$ was reached after a few cycles. Good convergence was 
achieved for values of $K_{\rm max}\approx 20-40$. Asymptotic values of 
$E_{\rm g.s.}$ were found by fitting the constants $C$ and $\alpha$ of 
the parametrization 
\begin{equation} 
E(K_{\rm max})=E_{\rm g.s.}+C\exp(-\alpha K_{\rm max}) 
\label{eq:conv} 
\end{equation} 
to values of $E(K_{\rm max})$ calculated for sufficiently high values of 
$K_{\rm max}$. The accuracy reached is better than 0.1 MeV in both the 
three-body and the four-body calculations reported here. 

The conversion width $\Gamma$ was then evaluated through the expression 
\begin{equation}
\Gamma = -2\, \langle \,\Psi_{\rm g.s.}\, | \, {\rm Im}\,V_{\eta N} 
\, | \, \Psi_{\rm g.s.} \, \rangle \;,
\label{eq:Gamma} 
\end{equation} 
where $V_{\eta N}$ sums over all pairwise $\eta N$ interactions. 
Since $|{\rm Im}\,V_{\eta N}|\ll |{\rm Re}\,V_{\eta N}|$, 
this is a reasonable approximation for the width.

\section{Results and discussion}
\label{sec:res}

Results of $\eta NN$ and $\eta NNN$ bound-state hyperspherical-based 
calculations for the GW $\eta N$ interaction, with Re~$a_{\eta N}$ almost 
1~fm, are given in this section. The weaker CS $\eta N$ interaction is found 
too weak to generate bound-state solutions. 
 
\subsection{$\eta NN$ calculations} 
\label{subsec:etaNN} 

No $I=0,\;J^{\pi}=1^-$ $\eta d$ bound state solution was found for the 
$\eta NN$ three-body system using the MN $NN$ potential \cite{Minn77} 
and the GW~\cite{GW05} $\eta N$ effective potential with a fixed strength 
$b(\delta\sqrt{s}=0)$, see Fig.~\ref{fig:Wycfit4.eps}, for either choice 
$\Lambda=2$ or 4~fm$^{-1}$ of the scale parameter under study. It was found 
that $b(\delta\sqrt{s}=0)$ in the GW model needs to be multiplied by 1.1 
for $\Lambda=4$~fm$^{-1}$ and by 1.3 for $\Lambda=2$~fm$^{-1}$ in order to 
generate a $1^-$ $\eta NN$ weakly bound state, with overall binding energy 
of $-$2.219 and $-$2.264 MeV, respectively, within three-body calculations 
that use a fixed $\eta N$ interaction strength $b(\delta\sqrt{s}=0)$. Recall 
that the MN deuteron binding energy is $E_d=-2.202$~MeV. There is no $\eta d$ 
bound state also in the $\eta N$ CS~\cite{CS13} model, judging by the CS/GW 
relative strengths of $b(\delta\sqrt{s})$. 

Given that the $\eta N$ interaction is too weak to bind the 
$I=0,\;J^{\pi}=1^-$ $\eta NN$ state in which the $^3S_1$ $NN$ (deuteron) 
core configuration is bound, the unbound $^1S_0$ $NN$ core configuration 
in the $I=1,\;J^{\pi}=0^-$ $\eta NN$ state certainly cannot support 
a three-body bound state. This holds so long as the $1^-$ state is unbound and 
also for a certain range of larger $\eta N$ potential strengths that make the 
$1^-$ bound. This situation is reminiscent of the $\Lambda NN$ system which 
is known to have {\it one} $I=0$ bound state in which the $\Lambda$ hyperon 
is bound to a deuteron core, but no $I=1$ $\Lambda NN$ bound state, see e.g. 
Ref.~\cite{GG14}. 

Our negative results rule out any $\eta d$ bound state, practically in 
all dynamical models of the $N^{\ast}(1535)$ resonance where the $\eta N$ 
interaction is coupled in, and are consistent with similar conclusions reached 
in Refs.~\cite{Deloff00,GP00,FA00,WG01,Gar03}. This holds also upon replacing 
the MN $NN$ interaction \cite{Minn77} by the AV4' $NN$ interaction \cite{WP02} 
in our $\eta NN$ calculations. In fact, somewhat larger $\eta N$ interaction 
multiplicative factors are then required to reach the onset of $\eta NN$ 
binding compared to those specified above. Applying the self-consistency 
requirement discussed in Sect.~\ref{sec:KLM} to the $\eta NN$ calculation, 
and recalling the decreased strength $b(\delta\sqrt{s})$ in the $\eta N$ 
subthreshold region, see Fig.~\ref{fig:Wycfit4.eps}, would only aggravate 
the failure to generate a three-body $\eta NN$ bound state. 

\subsection{$\eta NNN$ calculations} 
\label{subsec:etaNNN} 

Four-body $\eta NNN$ calculations were made using the MN \cite{Minn77} 
and the AV4' \cite{WP02} $NN$ potentials, and the GW \cite{GW05} and 
CS \cite{CS13} energy-dependent $\eta N$ potentials from Sect.~\ref{sec:v}. 
Based on $\eta ^3$H and $\eta ^3$He, and with the leading $3N$ configuration 
given by $I_N=S_N=\frac{1}{2}$ and $L_N^{\pi}=0^+$, the quantum numbers 
of the calculated $\eta NNN$ state are $I=S=\frac{1}{2}$, $L=0$ and 
$J^{\pi}={\frac{1}{2}}^-$. The $3N$ binding energy (disregarding the Coulomb 
interaction in the case of $^3$He) within our hyperspherical-basis calculation 
is $-$8.38~MeV for MN and $-$8.99~MeV for AV4'. Starting with the $\eta N$ GW 
model, with Re~$a_{\eta N}=0.96$~fm, and using the corresponding $v_{\eta N}$ 
from Sect.~\ref{sec:v} with {\it energy independent} threshold strength 
$b(\delta\sqrt{s}=0)$ for $\Lambda=4$~fm$^{-1}$, a four-body $\eta NNN$ 
bound state was found with $\eta$ separation energy $E_{\rm g.s.}^{\rm no~s.c.
}$ between 2 to 3~MeV, as listed in Table~\ref{tab:res}. We then applied 
a self consistency procedure by doing calculations with several given values 
of strength $b(\delta\sqrt{s})$, requiring that the expectation value 
$\langle\delta\sqrt{s}\rangle$ evaluated by Eq.~(\ref{eq:sqrt{s}}) from the 
obtained solution agrees with the input value of the subthreshold energy 
$\delta\sqrt{s}$ argument of the strength $b(\delta\sqrt{s})$ used in the 
calculation. This resulted in considerably reduced values of less than 1~MeV 
for the $\eta$ separation energy $E_{\eta~\rm sep.}^{\rm s.c.}$ which are 
listed in Table~\ref{tab:res}, together with the corresponding $\eta NNN$ 
binding energies $E_{\rm g.s.}^{\rm s.c.}$. Also listed in the table are the 
self consistent values $\delta\sqrt{s_{\rm s.c.}}$ and the self-consistency 
reduction factors $x_{\rm s.c.}\equiv b(\delta\sqrt{s_{\rm s.c.}})/b(\delta
\sqrt{s}=0)$. No $\eta NNN$ bound-state solutions were found using 
$v_{\eta N}^{\rm GW}$ self consistently for $\Lambda=2$~fm$^{-1}$. 

\begin{table}[htb] 
\begin{center} 
\caption{Results of $\eta NNN$ quasibound-state self-consistent calculations 
using the $\eta N$ model GW \cite{GW05}. Energies and widths are given in 
MeV.} 
\begin{tabular}{cccccccc} 
\hline 
$NN$ int. & E(NNN) & $E_{\rm g.s.}^{\rm no~s.c.}$ & 
$\delta\sqrt{s_{\rm s.c.}}$ & $x_{\rm s.c.}$ & $E_{\rm g.s.}^{\rm s.c.}$ & 
$E_{\eta~\rm sep.}^{\rm s.c.}$ & $\Gamma_{\rm g.s.}^{\rm s.c.}$ \\ 
\hline 
MN  & $-$8.38 & $-$11.26 & $-$13.52 & 0.914 & $-$9.33 & 0.95 & 13.52 \\  
AV4' & $-$8.99 & $-$11.33 & $-$15.83 & 0.895 & $-$9.03 & 0.04 & 15.75 \\  
\hline 
\end{tabular} 
\label{tab:res} 
\end{center} 
\end{table} 

\begin{figure}[htb] 
\begin{center} 
\includegraphics[width=0.8\textwidth]{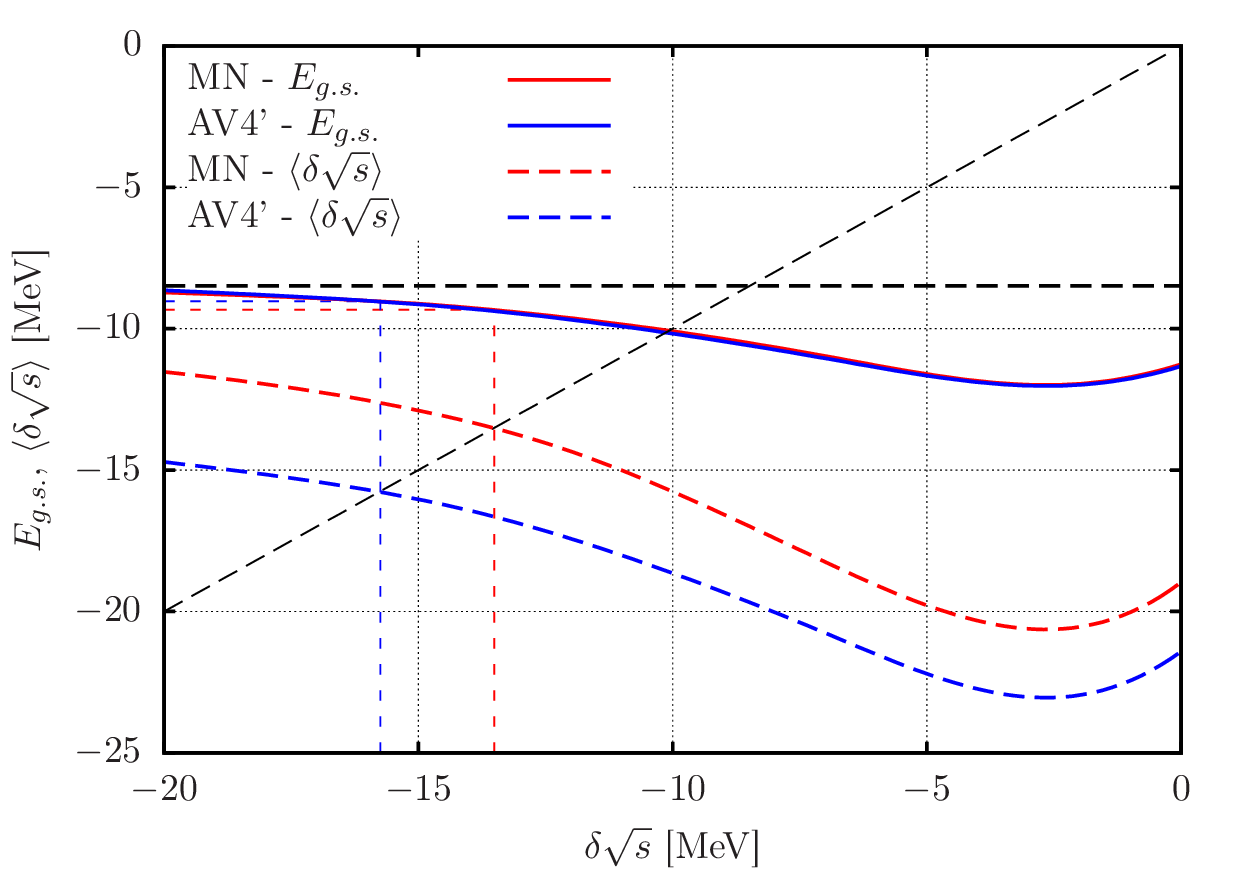} 
\caption{The $\eta NNN$ g.s. energy $E_{\rm g.s.}$ (solid curves) and 
the expectation value $\langle\delta\sqrt{s}\rangle$ (dashed curves) from 
Eq.~(\ref{eq:sqrt{s}}), calculated using the $NN$ potentials MN (red) and 
AV4' (blue), are shown as a function of the energy argument $\delta\sqrt{s}$ 
used for the input $v_{\eta N}^{\rm GW}$. The dashed horizontal line marks 
the $NNN$ ($^3$H) g.s. energy $-$8.48~MeV and the dashed diagonal line marks 
potentially self consistent solutions satisfying $\langle\delta\sqrt{s}\rangle
=\delta\sqrt{s}$. The dashed vertical lines mark the intersection of the 
dashed diagonal line with the $\langle\delta\sqrt{s}\rangle$ dashed curves, 
thereby fixing the self-consistent values $\delta\sqrt{s_{\rm s.c.}}$.}  
\label{fig:etaT_ds.ps} 
\end{center} 
\end{figure} 

In order to demonstrate how the self consistency procedure works we plotted 
in Fig.~\ref{fig:etaT_ds.ps} the $\eta NNN$ g.s. energy $E_{\rm g.s.}$ and 
expectation value $\langle\delta\sqrt{s}\rangle$, calculated as a function 
of the subthreshold energy $\delta\sqrt{s}$ argument of the input $\eta N$ 
potential $v_{\eta N}$ in both $NN$ potential models. The difference between 
the $E_{\rm g.s.}$ curves, using MN or AV4', is a fraction of MeV for any 
given input value $\delta\sqrt{s}$ and is hardly noticeable in the figure. 
The difference between the corresponding $\langle\delta\sqrt{s}\rangle$ curves 
amounts to a few MeV at each value of $\delta\sqrt{s}$ and is clearly visible 
in the figure, leading to self-consistency values $\delta\sqrt{s_{\rm s.c.}}$ 
which differ from each other by more than 2~MeV (marked by the dashed vertical  
lines). The corresponding self consistent values of $E_{\rm g.s.}$ are much 
closer to each other (marked by the thin dashed horizontal lines). The self 
consistency procedure is applied in the figure by looking for the intersection 
of the dashed diagonal line, locus of $\langle\delta\sqrt{s}\rangle=\delta
\sqrt{s}$, with each of the $\langle\delta\sqrt{s}\rangle$ dashed curves. 

Applying a similar self-consistency procedure to the weaker CS $\eta N$ 
interaction, rather than to the GW $\eta N$ interaction used above, 
no $\eta NNN$ bound state solution was found. With AV4' for the $NN$ 
interaction, this holds even upon using the threshold energy value in 
$v_{\eta N}^{\rm CS}$. With the MN $NN$ interaction and for the choice 
$\Lambda=4$~fm$^{-1}$, a bound-state solution is found for small values 
of the input energy $\delta\sqrt{s}$, disappearing at $-\delta\sqrt{s}
\approx 2.8$~MeV which is way below the minimum value of $-\delta\sqrt{s}$ 
required in the limit of $E_{\eta~\rm sep.}\to 0$. We conclude that the CS 
$\eta N$ interaction is too weak to provide self consistently $\eta NNN$ 
bound states. 

Finally, the $\eta NNN$ width $\Gamma_{\rm g.s.}^{\rm s.c.}\sim 15$~MeV 
listed in the last column of Table~\ref{tab:res} was calculated using 
Im~$b(\delta\sqrt{s_{\rm s.c.}})$ in forming the integrand Im~$V_{\eta N}$ in 
Eq.~(\ref{eq:Gamma}). This width is about three times larger than the widths 
evaluated self consistently using optical model methods across the periodic 
table within the $\eta N$ GW model \cite{FGM13}. Some explanation of this 
difference is offered noting that the magnitude of the downward energy shifts 
$\delta\sqrt{s_{\rm s.c.}}$ effective in those works is considerably larger 
by factors of two to three than the $\approx$15~MeV found in the present 
$\eta NNN$ calculations, reflecting the denser nuclear environment encountered 
by the $\eta$ meson as it becomes progressively more bound in the calculations 
of Ref.~\cite{FGM13}. Recalling the steady decrease of the $\eta N$ 
absorptivity Im~$F_{\eta N}$ in Fig.~\ref{fig:aEtaN1.eps} upon moving deeper 
into subthreshold energies, a factor of two to three difference could be 
anticipated in favor of relatively small $\eta$ widths in heavier nuclei.  



\section{Summary and outlook} 
\label{sec:concl} 

Precise hyperspherical-based few-body calculations were reported in this work 
to explore computationally whether or not $\eta$ mesons bind in light nuclei. 
To this end, complex energy-dependent local effective $\eta N$ potentials 
$v_{\eta N}$ were constructed, for subthreshold energies relevant to $\eta$ 
mesic nuclei, from coupled channel $\eta N$ scattering amplitudes in several 
models connected dynamically to the $N^{\ast}(1535)$ resonance. 
The scale dependence arising from working with an effective $v_{\eta N}$ 
was studied by using two representative values for the momentum scale, 
$\Lambda=2,4$~fm$^{-1}$. Noting that Im~$v_{\eta N}\ll\;$Re~$v_{\eta N}$, 
only the real part of $v_{\eta N}$ was used in the bound-state calculations, 
with a related error estimated as less than 0.2~MeV, added to an estimated 
0.1~MeV calculational error. The width of the bound state, making it into 
a quasibound state, was deduced from the expectation value of Im~$v_{\eta N}$ 
summed on all nucleons.  
 
No $\eta NN$ quasibound states were found for any of the two scale parameters 
chosen in models where the real part of the $\eta N$ threshold interaction 
satisfies Re~$a_{\eta N}\lesssim 1$~fm, in agreement with deductions made 
in several past few-body calculations of the $\eta d$ scattering length 
\cite{Deloff00,GP00,FA00,WG01,Gar03}. It is unlikely that the $\eta d$ system 
can reach binding upon increasing moderately the momentum scale parameter 
$\Lambda$. 

For $\eta NNN$, essentially the $\eta ^3$H and $\eta ^3$He isodoublet 
of $\eta$ mesic nuclei, a relatively broad and weakly bound state was 
found with $\eta$ separation energy of less than 1~MeV using the GW 
$\eta N$ interaction model \cite{GW05} where Re~$a_{\eta N}$ is almost 1~fm. 
This holds for the larger of the two values of momentum scale parameter, 
$\Lambda=4$~fm$^{-1}$, studied here, whereas no bound state was obtained 
upon using the smaller value of $\Lambda=2$~fm$^{-1}$. The energy dependence 
of $v_{\eta N}^{\rm GW}$, treated here within a self consistent procedure 
\cite{FGM13,CFGM14}, played an important role by reducing the calculated 
binding energy by about 2 MeV from that calculated upon using the $\eta N$ 
threshold energy value in $v_{\eta N}^{\rm GW}$. For such halo-like 
$\eta$-nuclear quasibound states, the neglect of Im~$v_{\eta N}$ in the 
bound-state calculation requires attention. In the case of the GW $\eta N$ 
effective interaction, we estimate the repulsion added by reinstating 
Im~$v_{\eta N}^{\rm GW}$ to second order to be roughly $\lesssim 0.2$~MeV, 
eliminating thereby the very weakly bound $\eta NNN$ state calculated here 
using the AV4' $NN$ potential, but not the weakly bound one calculated using 
the MN $NN$ potential. It is worth noting that the only other few-body 
$\eta NNN$ study known to us \cite{FA02} deduced from their calculated 
$\eta ^3$H scattering length that no quasibound state was likely. However, 
the strength of the $\eta N$ interaction tested in these calculations was 
limited to Re~$a_{\eta N}=0.75$~fm, short of our upper value of approximately 
1~fm.    

In conclusion, recalling the KW conjecture \cite{KW15} quoted in the 
Introduction, it is fair to say that the present few-body calculations 
support the conjecture's first and last items, namely that ``the $\eta d$ 
system is unbound" and ``that the $\eta ^3$He case is ambiguous". 
Accepting that the strength of the two-body $\eta N$ interaction indeed 
satisfies Re~$a_{\eta N}\lesssim 1$~fm, which is much too weak to bind the 
$\eta N$ system, a persistent theoretical ambiguity connected with choosing 
a physically admissible range of values for the $\eta N$ scale parameter 
$\Lambda$ is demonstrated by our few-body calculational results, particularly 
for the four-body $\eta NNN$ system. By choosing a considerably larger value 
of $\Lambda$ than done here one could bind solidly this system. To remove 
this ambiguity, many-body repulsive interactions involving the $\eta$ meson 
need to be derived and incorporated within few-body calculations. 

In future work we hope to extend our $\eta NNN$ calculations also by applying 
methods of complex scaling that should enable one to follow trajectories 
of $S$-matrix quasibound-state poles and look also for other types of poles 
such as virtual-state poles or resonance poles, all of which affect to some 
degree the threshold production features of $\eta$ mesons in association with 
$^3$He. Furthermore we hope to initiate a precise and realistic calculation 
of the $\eta NNNN$ system in order to test the middle item in the KW 
conjecture, namely that ``$\eta ^4$He is bound".

\section*{Acknowledgments} 

We thank Ji\v{r}\'{i} Mare\v{s} for carefully reading this manuscript and 
making useful remarks. This work was supported in part (NB) by the Israel 
Science Foundation grant 954/09, and in part (EF, AG) by the EU initiative 
FP7, Hadron-Physics3, under the SPHERE and LEANNIS cooperation programs.

\end{document}